\documentclass[aps,prl,citeautoscript,twocolumn,longbibliography,superscriptaddress]{revtex4-1}
\usepackage[T1]{fontenc}
\usepackage{latexsym}
\usepackage{graphicx} 
\usepackage{epstopdf}
\usepackage{amsmath}  
\usepackage{amssymb}
\usepackage{cases}
\usepackage{float}
\usepackage{tikz}
\usepackage{makecell}
\usepackage{bbold}
\usepackage{color}
\definecolor{aurometalsaurus}{rgb}{0.43, 0.5, 0.5}

\usepackage[breaklinks=true]{hyperref}
\usepackage{xcolor}

\newcommand{\apatite}{Pb\textsubscript{10-$x$}Cu\textsubscript{$x$}(PO\textsubscript{4})\textsubscript{6}O }

\begin{document}

\title{High Temperature Superconductivity with Strong Correlations and Disorder:\\
Possible Relevance to Cu-doped Apatite}

\author{Maciej Fidrysiak}
\email{maciej.fidrysiak@uj.edu.pl}
\affiliation{Institute of Theoretical Physics, Jagiellonian University, ul. \L{}ojasiewicza 11, PL-30-348 Krak\'ow, Poland}
\author{Andrzej P. K\k{a}dzielawa}
\email{andrzej.kadzielawa@uj.edu.pl}
\affiliation{Institute of Theoretical Physics, Jagiellonian University, ul. \L{}ojasiewicza 11, PL-30-348 Krak\'ow, Poland}
\author{J\'ozef Spa\l{}ek}
\email{jozef.spalek@uj.edu.pl}
\affiliation{Institute of Theoretical Physics, Jagiellonian University, ul. \L{}ojasiewicza 11, PL-30-348 Krak\'ow, Poland}

\begin{abstract}
We examine the properties of topological strongly correlated superconductor with bond disorder on triangular lattice and demonstrate that our theoretical ($t$-$J$-$U$) model exhibits some unique features of the Cu-doped apatite \apatite. Namely, the paired state appears only for carrier concentration $0.8 \lesssim n < 1$ per Cu and is followed by a close-by phase separation into the superconducting and Mott insulating parts. Furthermore, a moderate amount of the bond disorder ($\Delta t /  t \lesssim 20 \%$) does not alter essentially the topology with robust Chern number $C=2$ which diminishes beyond that limit. A room-temperature superconductivity is attainable only for the  exchange to hopping ratio $J/|t| \ge 1$ if one takes the bare bandwidth suggested by current DFT calculations. The admixture of $s$-wave pairing component is induced by the disorder. The results have been obtained within statistically consistent variational approximation (SGA).
\end{abstract}

\maketitle

\emph{Motivation.}---The claimed \cite{LeeArxiv:2307.12008,LeeArxiv:2307.12037} and partly confirmed \cite{LeeArxiv:2308.15016} room-temperature superconductivity at ambient pressure in Cu-doped apatite \apatite  with $x \approx 1$, created an unprecedented activity mixed with persistent hopes of achieving room temperature superconducting technology and quantum computer devices. From theoretical side, the principal question is whether it is possible to extend the degree of relevancy achieved by the Bardeen, Cooper, and Schrieffer (BCS) \cite{BardeenPhysRevB1957} and Eliashberg \cite{EliashbergJETP1960,EliashbergJETP1961} approaches in order to make them similarly effective when applied to novel hydrogen-rich systems \cite{LiuPNAS2017,PengPhysRevLett2017} on one side, and to the high-temperature cuprates \cite{AndersonScience1987,OgataRepProgPhys2008,LeeRevModPhys2006,SpalekPhysRep2022} and the just discovered strongly correlated systems on the other.

\begin{figure}
  \centering
	\includegraphics[width=1\linewidth]{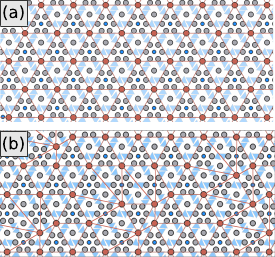}
  \caption{Illustration of triangular lattice with copper (red) and oxygen (blue), emulating pristine (undistorted) system and that with randomly distorted bond arrangements [(a) and (b), respectively]. Grey balls represent Pb ions. The $\mathrm{PO_4}$ groups are marked in the background (shaded tetrahedrons).}
	\label{fig:one}
\end{figure}

Here we show that solving the $t$-$J$-$U$ model, applied earlier to the cuprates \cite{SpalekPhysRep2022,SpalekPhysRevB2017,FidrysiakJPCM2018} and appropriately reformulated, can provide a number of features  observed very recently \cite{LeeArxiv:2307.12008,LeeArxiv:2307.12037,LeeArxiv:2308.15016} under stringent conditions. To discuss those in detail, we start by noting that the Cu-doped apatite system in question has a well defined triangular substructure of Cu atoms surrounded by free oxygen and strongly bound PO\textsubscript{4} groups. The substructure is illustrated in Fig.~\ref{fig:one}, where we have shown the arrangement without (a) and with (b) Cu substitution disorder. We assume that since the Cu-Cu distance is quite large ($\sim 10$~\AA) a narrow (flat) band may be formed due to the intersite hybridization of Cu (red balls) with surrounding ions, as obtained very recently \cite{KurletoArxiv:2308.00698,SiArxiv:2308.00676}. Such a model is thus interesting in its own right \cite{ChakrabortyNPJQuantMater2022}, as discussed below.

In the second part, we discuss the relevance of our results to the properties of Cu-doped apatite. We analyze unique scenario, emerging when the topology, strong correlations, and atomic disorder are all combined together. Detailed of the model and theoretical scheme are provided below.

\begin{figure}[h]
	\centering
	\includegraphics[width=1\linewidth]{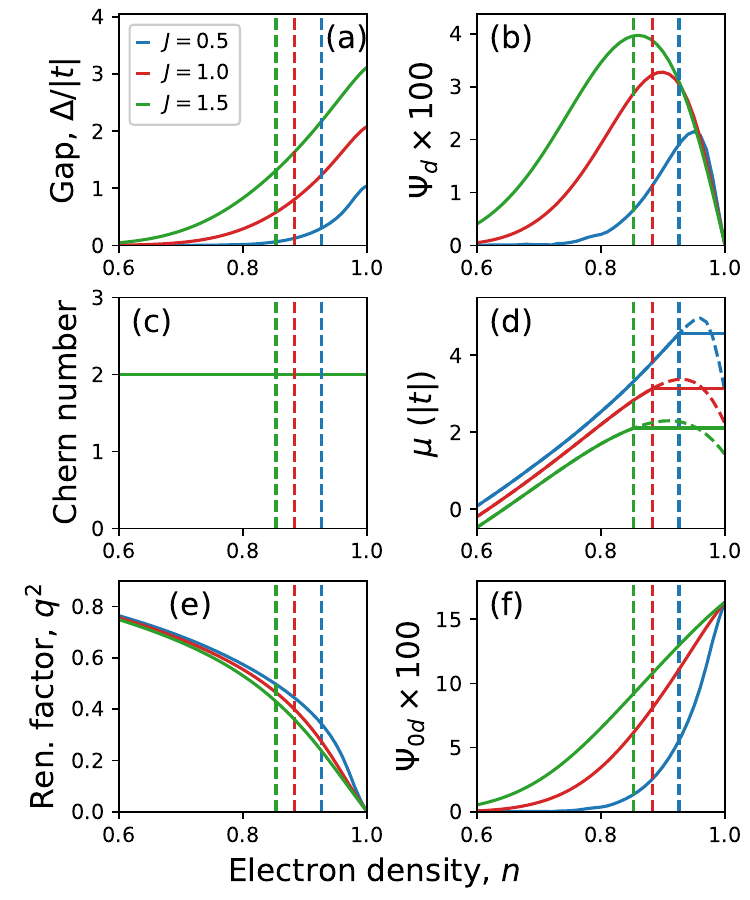}
	\caption{Basic properties for pristine lattice case, all as a function of electron density per Cu atom and for three values of exchange to hopping ratio $J/|t| = 0.5, 1.0$, and $1.5$. The remaining parameters are defined in the text. The vertical dashed lines mark the phase separation line between topological $d+id$ superconductor and Mott insulator. (a) Energy gap in the quasiparticle spectrum (SC state is nodeless). (b) True $d+id$ SC order parameter amplitude (evaluated with correlated wave function), showing dome-like shape in the range $n \in (0.8, 1.0)$ and disappearing in the Mott state. (c) Robustness of the Chern number, $C=2$, as obtained numerically in the SC regime. (d) Chemical potential with $\partial \mu / \partial n < 0$ regime signaling phase separation. (e) Band narrowing factor driven by strong correlations. The $q^2 = 0$ ($n=1$) point marks the Mott insulating state. (f) The uncorrelated SC amplitude exhibiting a pseudogap behavior [note similarity between panels (a) and (f)].}
	\label{fig:two}
\end{figure}

\begin{figure*}
  \centering
  \includegraphics[width=1\linewidth]{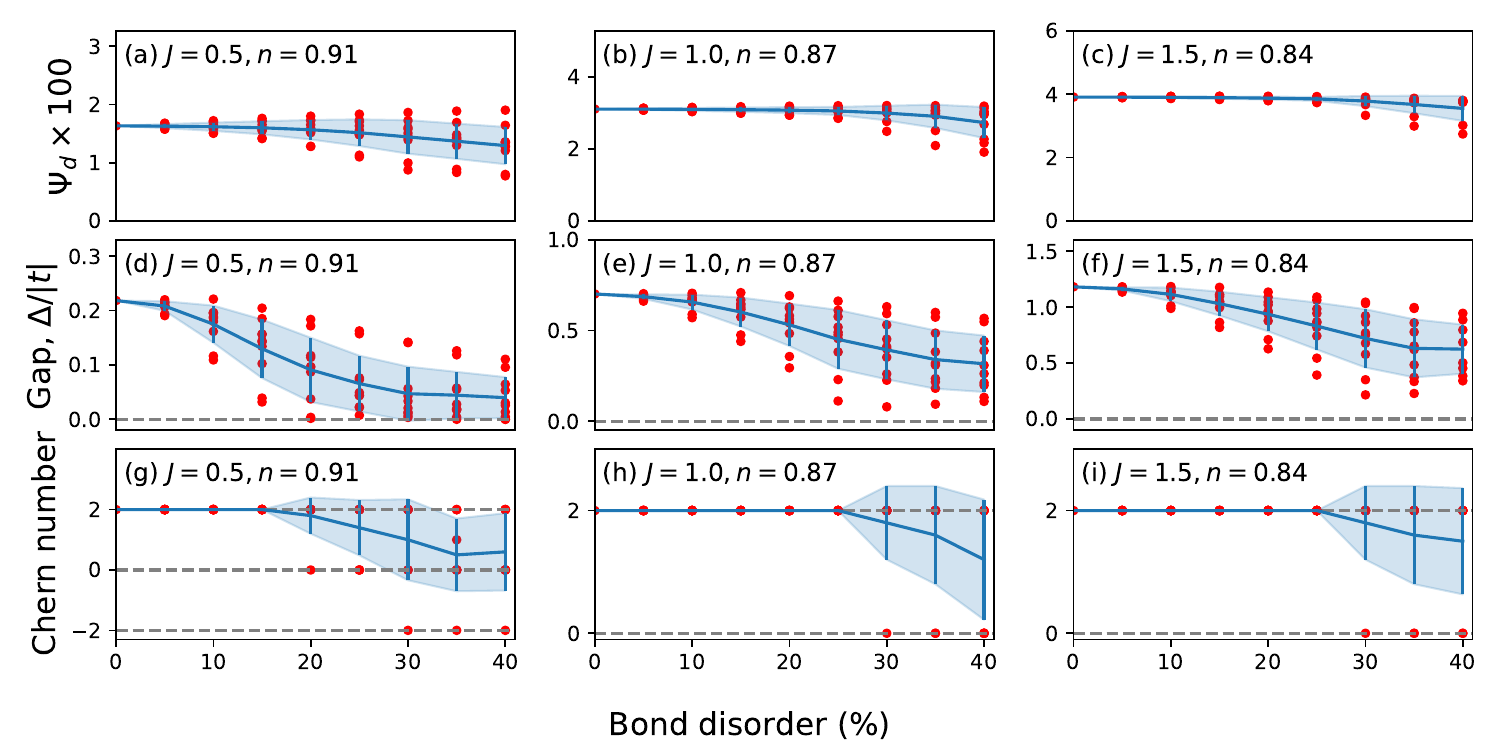}
  \caption{Evolution of the superconducting-state properties as a function of bond disorder magnitude $ |\Delta t| / |t|$ for thee values of antiferromagnetic exchange, $J/|t| = 0.5, 1.0, 1.5$, and fixed electron densities $n = 0.91, 0.87$, and $0.84$, respectively. The values of $n$ have been selected just outside the phase-separation regime. The specified quantities are: (a)-(c) the $d+id$ SC order parameter $\Psi_d$, (d)-(f) energy gap $\Delta$, and (g)-(i) Chern number. The data scattering for $8$-$10$ random realizations used for sampling is marked by red points. Solid lines and shaded represent the averaged quantities and the mean square deviations, respectively. The disorder up to $\sim 30\%$ does not alter essentially the data trend for large $J$. For $J/|t| = 0.5$ the energy gap diminishes substantially with increasing disorder and topological superconductivity is suppressed rapidly.}
  \label{fig:disorder_main}
\end{figure*}

\begin{figure}
  \centering
  \includegraphics[width=1\linewidth]{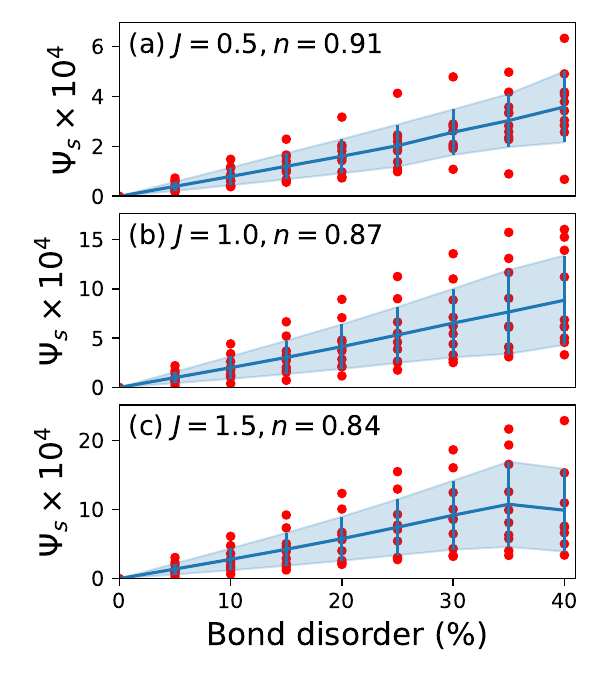}
  \caption{Magnitude of the $s$-wave order parameter, $\Psi_s$, for the microscopic parameters the same as those used to generate Fig.~\ref{fig:disorder_main}. Note that the magnitude of $\Psi_s$ is much smaller than the dominant $d+id$ amplitude, $\Psi_d$.}
  \label{fig:disorder_swave}
\end{figure}

\begin{figure}
  \centering
  \includegraphics[width=1\linewidth]{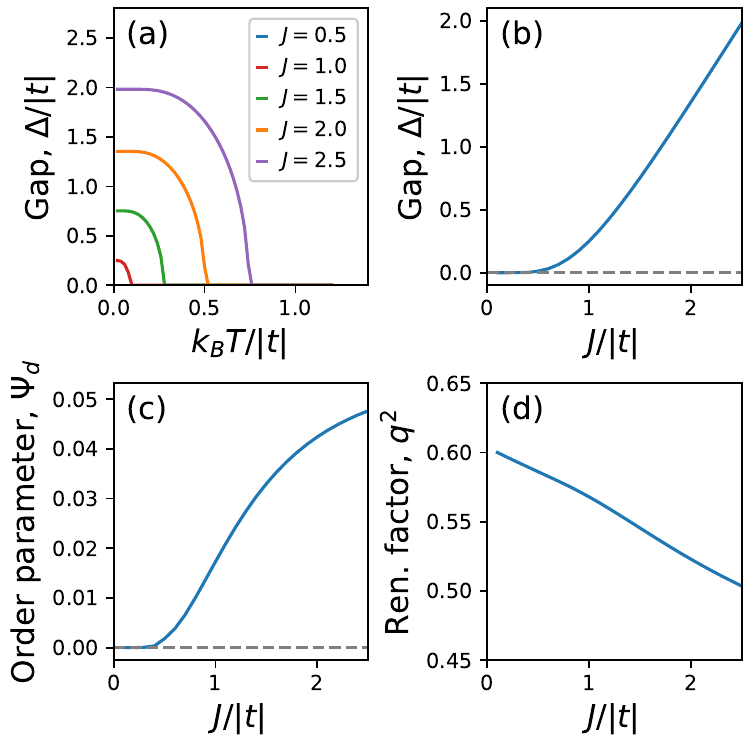}
  \caption{Scaling of SC properties with temperature and exchange coupling without disorder for $n = 0.79$. (a) Temperature dependence of the energy gap. (b)-(d) Gap $\Delta$, $d$-wave order parameter $\Psi_d$, and band renormalization-factor $q^2$, all as a function of relative exchange magnitude, $J/|t|$. Note that $\Psi_d$ tends to saturate at large $J$, so increase of $\Delta$ is driven by the corresponding increase of $J$, since $\Delta \sim J$.}
  \label{fig:scaling}
\end{figure}

\emph{Model.}---We employ the $t$-$J$-$U$ model \cite{SpalekPhysRevB2017} on triangular lattice, given by the Hamiltonian

\begin{align}
  \hat{\mathcal{H}} = \sum_{i \neq j, \sigma} t_{ij} \hat{a}_{i\sigma}^\dagger \hat{a}_{j\sigma} + J \sum_{\langle i, j\rangle} \hat{\mathbf{S}}_i \hat{\mathbf{S}}_j + U \sum_i \hat{n}_{i\uparrow} \hat{n}_{i\downarrow},
\end{align}

\noindent
where $\hat{a}_{i\sigma}$ ($\hat{a}_{i\sigma}^\dagger$) annihilate (create) electrons on the Cu sites, $\hat{\mathbf{S}}_i$ are spin operators, and $\hat{n}_{i\sigma} \equiv \hat{a}^\dagger_{i\sigma} \hat{a}_{i\sigma}$. We include nearest- and next-nearest-neighbor hopping integrals, $t < 0$ and $t^\prime = 0.4 |t|$, respectively. The on-site Coulomb repulsion is set to $U = 16 |t|$, and $J$ varies in the range $0.5$-$1.5 |t|$. If not stated otherwise, temperature is $k_B T = 10^{-6} |t|$. Parenthetically, the present one-band model of high-temperature superconductivity may be regarded as a generic effective Hamiltonian for the three-orbital $d$-$p$ model in the same manner as one-band model represents canonical case for its three-band correspondent for the cuprates \cite{ZegrodnikPhysRevB2019}. Extended discussion of the model selection and its parameters is deferred to Supplementary Material \cite{SupplementaryMaterial}.

\emph{Results.}---In Fig.~\ref{fig:two}, composing the panels (a)-(f), we present the basic properties in the nodeless superconducting (SC) state with the gap of $d_{x^2-y^2}+i d_{xy}$ character that is compatible with the triangular lattice symmetry, all as a function of electron density $n$ per Cu atom. Explicitly, panel (a) represents the gap in quasiparticle spectrum, in units of the nearest-neighbor (n.n.) hopping integral $|t|$ and for three different values of n.n. antiferromagnetic exchange $J$ (in the same units). This result is in contrast with that displayed in figure (b), showing the true $d+id$ SC order parameter

\begin{align}
  \label{eq:order_parameter}
  \Psi_{d} \equiv  \left| \frac{1}{3N} \sum_{\substack{n = 1, 2, 3 \\ i \in \text{lattice}}} \mathrm{e}^{-i \frac{2\pi}{3} \cdot n} \langle\hat{a}_{i + \hat{\delta}_n, \downarrow} \hat{a}_{i,\uparrow} \rangle \right|,  
\end{align}

\noindent
where $i$ runs over all lattice sites and $\hat{\delta}_1 = (1, 0)$, $\hat{\delta}_2 = (\frac{1}{2}, \frac{\sqrt{3}}{2})$, $\hat{\delta}_3 = (-\frac{1}{2}, \frac{\sqrt{3}}{2})$ are vector connecting nearest-neighbor sites (in the lattice units), and $N$ denotes the number of copper atoms. Note that it is sufficient to incorporate only summation over three (out of six) n.n. SC amplitudes, as for spin-singlet pairing  $\sum_i \langle\hat{a}_{i + \hat{\delta}_n, \downarrow} \hat{a}_{i,\uparrow}\rangle = \sum_i \langle\hat{a}_{i - \hat{\delta}_n, \downarrow} \hat{a}_{i,\uparrow}\rangle$, which we have also verified numerically. The expectation value in Eq.~\eqref{eq:order_parameter} is evaluated in the fully relaxed variational state, within the \textbf{S}tatistically-\textbf{C}onsistent \textbf{G}utzwiller (SGA) approximation with local correlators \cite{SupplementaryMaterial,SpalekPhysRep2022}. The order parameter $\Psi_d$ exhibits typical dome-like behavior present for the cuprates, except that here a nonzero solution appears in a relatively narrow carrier concentration interval. Figure~\ref{fig:disorder_main}(c) illustrates the robustness of the Chern number $C=2$ to doping, obtained numerically by efficient Brillouin-zone triangulation method \cite{FukuiJPSJ2005}. Figure \ref{fig:disorder_main}(d) shows the nonmonotonic dependence of the chemical potential $\mu$ on electron density ($\tfrac{\partial \mu}{\partial n} < 0$ for $n \approx 1$), signaling phase separation. This circumstance may explain, in a natural manner, the reported difficulties with detecting a clear zero-resistivity state and small volume of the SC phase \cite{HouArxiv:2308.01192}. The vertical dashed lines mark phase separation boundaries obtained by the Maxwell construction, as detailed in panel (d). Finally, in (e) and (f) the correlation-induced band narrowing factor $q^2$ and the uncorrelated $d+id$ order parameter $\Psi_{0d}$ (obtained by disregarding correlations in the variational state) are shown. The quasiparticle bandwidth vanishes ($q^2 = 0$) in the Mott insulating state ($n=1$). Also, the $\Psi_{0d}$ may be interprated as a pseudogap within the gossamer SC scenario \cite{LauglinPhilMag2006,SpalekPhysRep2022}. In summary, the above results for pristine (undistorted) situation are similar to those obtained for the cuprates, except for nontrivial topology ($C=2$) and nodeless SC gap. Fully-gapped SC state is consistent with first experimental results for \apatite \cite{LeeArxiv:2307.12008,LeeArxiv:2307.12037}.

A physical interpretation of the presented results is in place here. Band structure calculations \cite{KurletoArxiv:2308.00698,SiArxiv:2308.00676,GriffinArixiv:2307.16892} suggest that the bare bandwidth $W$ of the relevant electronic states near Fermi energy is 0.1-0.3 eV. Taking $W = 9 |t| = 0.2 \, \mathrm{eV}$ for the triangular lattice and assuming value of SC energy gap $\Delta/|t| \approx 1$ (cf. Fig.~\ref{fig:two}(a) for $J/|t| = 1.0$-$1.5$ at the uniform SC stability edge), one obtains the gap magnitude in physical units $\Delta \approx 0.022 \, \mathrm{eV} \approx 260 \, \mathrm{K}$, approximately the same as the value of exchange integral. In order to obtain the room temperature $\Delta$ value the order of $J$ magnitude must be  comparable to that of the hopping integral. The principal question is then if such situation with such relatively large $J$ is attainable in this case. A direct answer is that close to the Mott transition, the bare bandwidth $W \sim 0.2\,\mathrm{eV}$, whereas renormalized quasiparticle bandwidth $\sim q^2 W$ is reduced, but still exceeds by far the superexchange $\sim  20\,\mathrm{meV}$ (even though $J \sim |t|$).

We now proceed to a more realistic situation, depicted in Fig.~\ref{fig:one}(b), where Cu positions are subjected to chemical disorder due to lead site substitution. Physically, this sort of disorder affects relative distances between Cu sites and may be modeled as the variation of hopping integrals, $t_{ij} \rightarrow t_{ij} \cdot (1 + \eta_{ij})$, with $\eta_{ij}$ being selected randomly (we assume uniform distribution on the interval $[-\eta_\mathrm{max}, \eta_\mathrm{max}]$, where $\eta_\mathrm{max}$ is hereafter dubbed ``bond disorder'' parameter). We carry out simulations for $8$-$10$ random realization selections on $200 \times 200$ lattice of $2 \times 2$ supercells (within which hopping integrals are randomized). Subsequently, the disorder-averaged physical quantities and mean-square variation corresponding to those in Fig.~\ref{fig:two} are evaluated. The results are summarized in the panel composing Fig.~\ref{fig:disorder_main}. Explicitly, each row is plotted for $J/|t| = 0.5, 1.0$, and $1.5$ (detailed inside the panels), and for specified values of $n$ selected to stay clear of phase separation for $\eta_\mathrm{max} \rightarrow 0$ (cf. Fig.~\ref{fig:two}). The first row (a)-(c) shows robustness of the $d$-wave order parameter $\Psi_d$ against disorder. The second and third rows indicate a more substantial reduction of both the energy gap magnitude $\Delta$ and averaged Chern number, above $\eta_\mathrm{max} \sim 20 \%$. Note that Chern numbers for each individual disorder realization are integers, but their average is not; its reduction signals breakdown of topological state. Noteworthy, disorder substantially impacts the SC gap for $J/|t| = 0.5$, but both SC gap and topology remain robust even up to $\eta_\mathrm{max} \sim 30\%$ for larger values of $|J|$.

Another nontrivial aspect of Cu positional disorder is mixing between order parameters of different symmetries, as a consequence of reduced symmetry at the Cu lattice site. This is demonstrated in Fig.~\ref{fig:disorder_swave}, where the average (solid line), individual solutions (red points), and the standard deviation (shaded areas) are shown for the (extended) $s$-wave order parameter

\begin{align}
  \label{eq:order_parameter_swave}
  \Psi_{s} \equiv  \left| \frac{1}{3N} \sum_{\substack{n = 1, 2, 3 \\ i \in \text{lattice}}} \langle\hat{a}_{i + \hat{\delta}_n, \downarrow} \hat{a}_{i,\uparrow} \rangle \right|.
\end{align}

\noindent
Note that its magnitude is much smaller from that for $d$-wave symmetry, but systematically increases with increasing bond disorder. In addition to the pairing component defined by Eq.~\eqref{eq:order_parameter_swave}, we also observe local pairing amplitudes  $\langle\hat{a}_{i\downarrow} \hat{a}_{i\uparrow}\rangle \neq 0$. Analogous mixing of $s$-wave and $d$-wave order parameters has been reported for coexistent charge-density wave and nematic phase with SC state in high-temperature copper-oxide superconductors \cite{ZegrodnikEPJB2020,ZegrodnikPhysRevB2018}.

Finally, to examine the asymptotic behavior superconductivity and assess the trends, we have plotted in Fig.~\ref{fig:scaling} the following quantities: (a) the temperature behavior of the energy gap $\Delta$, which within SGA exhibits a mean-field-like behavior; (b) its dependence on $J$, which for $J/|t| \gtrsim 1$ becomes  linear, speaking in favor of local (practically n.n.) pair formation; (c) the $d$-wave order parameter $\Psi_d$ and (d) the narrowing factor $q^2$ showing the effect of saturation, which also supports the local pairing scenario. Based on these points, it is intriguing to inquire whether thinking of effective dense Bose-Einstein condensate formation for $J > |t|$ is possible. The argument with $\tfrac{\partial \mu}{\partial n} < 0$ speaks rather for the Mott insulating state prelevance instead, with the formation of n.n. spin singlets, perhaps of quantum spin-liquid nature.

\emph{Discussion.}---We have presented results obtained in SGA for strongy correlated electrons ($U/|t| = 16$) on a taken here substantially distorted triangular lattice. A number of results provides attractive features of the paired state \cite{LeeArxiv:2307.12008,LeeArxiv:2307.12037}. Namely,

\begin{enumerate}
	\item[(i)] the paired $\Psi_d$ state with high transition temperatures appears only close to half filling $1 > n \gtrsim 0.8$,
	\item[(ii)] phase separation is close-by to the paired state (in contrast to square-lattice model, relevant to high-$T_c$ cuprates),
	\item[(iii)] appearance of high-temperature SC state is possible for $J \gtrsim |t|$, i.e., in the superexchange dominated area ($J\approx 20 \, \mathrm{meV}$) which may be feasible in view of the circumstance that $|t| \gtrsim 20 \, \mathrm{meV}$,
	\item[(iv)] superconducting order parameter $\Psi_d$ vanishes in the Mott insulating state.	
\end{enumerate}

\emph{Outlook.}---Having the model which contains a number of attractive features, its natural extensions may make it realistic. This would include, first of all, an extension to the anisotropic three-dimensional situation. The second factor would be then to formulate a more realistic three-dimensional model of the Mott insulator or noncollinear antiferromagnetic state. In two dimensions, the reference state may be that of frustrated quantum spin liquid. Furthermore, the disorder-induced appearance of the SC $s$-wave solution opens up a possibility of including electron-phonon interaction on the same footing with the exchange, albeit in the correlated state (i.e., with renormalized microscopic parameters). Parenthetically, such a procedure seems possible to be included within our SGA approach, although quite cumbersome to implement for this complex lattice as it may possess soft modes. Finally, the phase-separated Mott-insulating islands and SC may compose natural Josephson junctions between crystallites in polycrystalline material. Such a circumstance may result in observable quantum coherence effects in inhomogeneous samples, as previously reported, e.g., for twisted-bilayer graphene \cite{CaoNature2018}.

\section{Acknowledgements}
\label{sec:acknowledgements}
The authors are very grateful to Dariusz Kaczorowski, Krzysztof Rogacki, and Tomasz Klimczuk for discussions concerning structural and superconducting properties of the Cu-doped apatite and related systems. Helpful discussions with Sergiu Arapan are also appreciated. The work was supported by Narodowe Centrum Nauki (NCN), Grant No. UMO-2021/41/B/ST3/04070.

\clearpage
  

\renewcommand{\thepage}{\arabic{chapter}.\arabic{page}} 
\renewcommand{\thesection}{\arabic{chapter}.\arabic{section}}  
\renewcommand{\thetable}{\arabic{chapter}.\arabic{table}}  
\renewcommand{\thefigure}{\arabic{chapter}.\arabic{figure}}
\renewcommand{\thepage}{S\arabic{page}} 
\renewcommand{\thesection}{S\arabic{section}}  
\renewcommand{\thetable}{S\arabic{table}}  
\renewcommand{\thefigure}{S\arabic{figure}}
\renewcommand{\figurename}{Figure}
\renewcommand*{\citenumfont}[1]{S#1}
\renewcommand*{\bibnumfmt}[1]{[S#1]}

\setcounter{figure}{0}
\setcounter{equation}{0}
\setcounter{page}{1}

\section{Supplementary Material}

In this Supplementary Material we discuss, first of all, details of our model and methods, as well as elaborate on selected aspects of our analysis within the statistically-consistent variational approximation (SGA).

\section{Model}

We have chosen one-band $t$-$J$-$U$ model (cf. Eq.~(1) of the main text) of strongly correlated electrons (for review, see \cite{SpalekPhysRep2022Suppl}) which represents an extended $t$-$J$ model, and incorporates both the exchange interaction $J$ and the Hubbard repulsion $U$ on the same footing. There are two principal reasons for that. The first of them is formal. Namely, including $U$ and subsequently considering the strong correlation limit $W/U \ll 1$ (with $W$ being bare bandwidth), we can omit formally introducing difficult to handle projected fermion-operator representation necessary in precise derivation of the $t$-$J$ model, as discussed originally in \cite{ChaoJPhysC1977Suppl} (see also \cite{SpalekPhysRep2022Suppl}). In the $t$-$J$-$U$ formulation, this projection occurs automatically in the $U \rightarrow \infty$ limit as then the doubly-occupied sites are suppressed. Second, superexchange may take place through a number of channels, e.g., involving also oxygen ions from $\mathrm{PO_4}$ groups, as well as the lone oxygen present in the \apatite structure, that position themselves in between the Cu sites. In effect, the relation between the underlying Hubbard-type Hamiltonian and $J$ obtained within the canonical perturbation theory, $J \sim 4t^2/U$ ($t$ denotes the hopping integral between the Cu sites), does not hold here. Indeed, in \apatite there is a substantial number of oxygen sites in the Cu neighborhood and this circumstance may help in achieving the value of $J$ of comparable magnitude to that of $|t|$ or even exceeding it. This scenario is also favored by unusually narrow bare bands ($0.1$-$0.3 \, \mathrm{eV}$) formed due to substitution of Pb by Cu, as evidenced by most recent DFT calculations. In absolute units, the value of superexchange $J$ required to support room temperature superconductivity is still by at least a factor of two smaller from that in high-temperature copper-oxide superconductors, even though its relative value $J/|t|$ is large due to smallness of $|t|$. Of course, these qualitative statements require a more quantitative analysis, including identification and summation of the contributions from dominant superexchange pathways, which may not be easy.

\section{Copper sublattice with bond disorder}

\begin{figure*}
  \centering
	\includegraphics[width=1\linewidth]{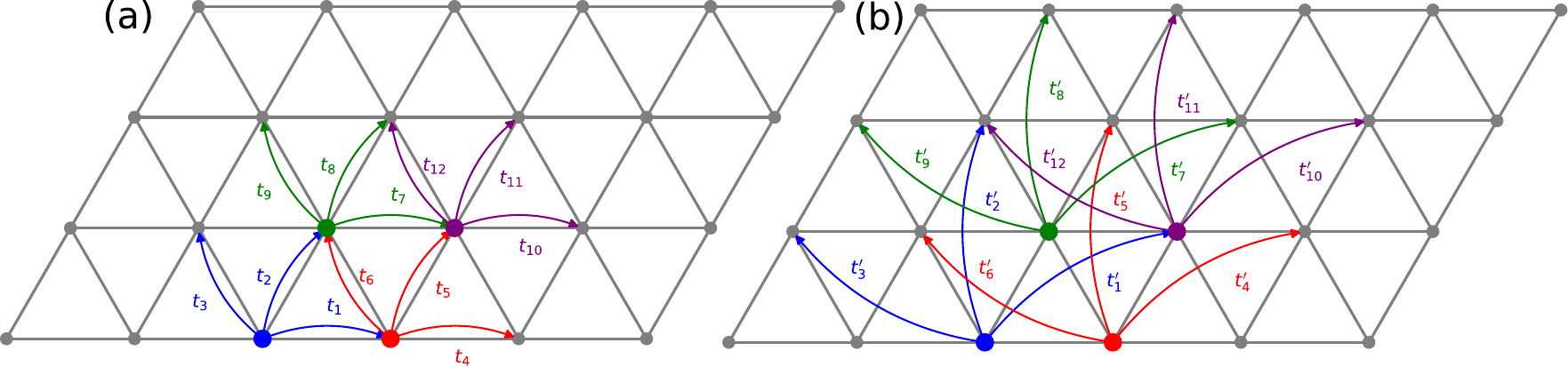}
  \caption{Illustration of the $2 \times 2$ Cu supercell employed to analyze disorder effects. Blue, red, green, and purple sites mark four independent Cu positions. Panels (a) and (b) show independent nearest- and next-nearest-neighbor hopping integrals, respectively (link colors have been chosen the same as those of the origin site for clarity). The magnitudes of both $t_i$ and $t^\prime_i$ for $i = 1, \ldots, 12$ are randomized according to the procedure described in the text. }
	\label{fig:lattice_hopping}
\end{figure*}

To study the pristine case (Fig.~1(a) of the main text) we use triangular $400 \times 400$ lattice with periodic boundary conditions ($1.6 \times 10^5$ sites). In order to account for the physics of positional Cu ion disorder [cf. Fig.~1(b) of the main text], we employ the following construction. The pristine lattice is divided into supercells containing four ions each, forming a $200 \times 200$ superlattice so that the total number of atoms is preserved. In Fig.~\ref{fig:lattice_hopping}, four independent Cu sites are marked in blue, red, green, and purple circles. Panels (a) and (b) of Fig.~\ref{fig:lattice_hopping}) show the independent nearest- and next-nearest-neighbor hopping integrals, respectively. Random configurations for disorder sampling are generated by setting $t_i \rightarrow  t_i (1 + \eta_i)$ and $t^\prime_i  \rightarrow t_i^\prime (1 + \eta_i^\prime)$, where index $i = 1, \ldots, 16$ and $\eta_i$, $\eta^\prime_i$ are pseudo-random numbers generated from uniform distribution on the interval $[-\eta_\mathrm{max}, \eta_\mathrm{max}]$. The bond disorder parameter, $\eta_\mathrm{max}$, is used as a dimensionless proxy measure of disorder induced by inhomogeneous Pb substitution as this sort of spatial displacements affect the hopping integrals between Cu centers. 

The disorder-averaged quantities, displayed in Figs. 2 and 3 of the main text are calculated based on $10$ randomized configurations of hopping integrals, except for the case $J = 1.5 |t|$ and $\eta_\mathrm{max} = 40 \%$, where the convergence could not be achieved for two random seeds, regardless of our solver configuration. We consider this behavior as physical manifestation of the proximity to phase separation regime between superconducting and Mott insulating states. Namely, sufficiently large disorder is capable of driving the system toward this instability. The mean-square error estimate for a given physical quantity $x$ is estimated as $\sigma(x) = \sqrt{\frac{1}{M} \sum_i (\bar{x} - x_i)^2}$, where summation runs over realizations of randomness, $M$ denotes number of sampled configurations, and $\bar{x} \equiv \frac{1}{M} \sum_i x_i$.

\section{Method}

The method we use is the variational approach based on modified Gutzwiller approach, the \textbf{S}tatistically-\textbf{C}onsistent \textbf{G}utzwiller (SGA) method \cite{SpalekPhysRep2022Suppl,BunemannEPL2012Suppl,JedrakArxiv:1008.0021Suppl}. In the lowest-order approximation, in which the microscopic parameters, $t$ and $J$, are renormalized by the correlations by the amount dependent on electron density, $n$. Additionally, a number of constraints are introduced to assure the statistical consistency of the results \cite{JedrakArxiv:1008.0021Suppl}. In effect, the method removes the defect of the Gutzwiller approach and thus the results obtained either by variational minimization or by solving self-consistent equations, coincide \cite{JedrakArxiv:1008.0021Suppl}. In essence, this procedure amounts to fulfilling the Bololiubov principle any mean-field-type or other approximation should obey. The leading order SGA results are comparable to those coming from the slave-boson approach, but without introducing the auxiliary Bose fields that may undergo a spurious Bose-Einstein condensation at the man-field level \cite{JedrakArxiv:1008.0021Suppl}.

Here we briefly outline used by us nonstandard variant of SGA latter that incorporates on-site superconducting pairing on equal footing with intersite amplitudes. This development is essential in the context of considered here distorted Cu lattice, since locally reduced symmetry generally leads to admixture of $s$-wave component to the dominant $d$-wave amplitude (cf. discussion in the main text). Similar symmetry-mixing effects have been recently noted in coexistent charge-density-wave/nematic and superconducting phase for models of high-temperature copper-oxide superconductors \cite{ZegrodnikEPJB2020Suppl,ZegrodnikPhysRevB2018Suppl}.

The starting point for the approach is variational wave function, $|\Psi_\mathrm{var}\rangle \equiv \hat{P} |\Psi_0\rangle$, where $|\Psi_0\rangle$ is an uncorrelated (Slater-determinant-type) state and $\hat{P}$ is an operator introducing correlations into $|\Psi_\mathrm{var}\rangle$. Note that $|\Psi_0\rangle$ is not a plain Fermi-sea state, but it encodes information about broken symmetries and topology. We take $\hat{P} \equiv \prod_i \hat{P}_i$ as a product of local correlators

\begin{align}
  \label{eq:correlator}
  \hat{P}_i \equiv \lambda_{i, 0} |0\rangle_i{}_i\langle{0}| + \sum_\sigma \lambda_{i, \sigma} |\sigma\rangle_i{}_i\langle{\sigma}| + \lambda_{i, d} |d\rangle_i{}_i\langle{d}|.
\end{align}

\noindent
In Eq.~\eqref{eq:correlator}, the Dirac-notation projection operators onto the local basis states ($|{0}\rangle_i$, $|{\uparrow}\rangle_i$, $|{\downarrow}\rangle_i$, and $|{d}\rangle_i \equiv |{\uparrow\downarrow\rangle_i}$) are weighted by variational parameters ($\lambda_{i, 0}$, $\lambda_{i, \sigma}$, and $\lambda_{i, d}$) to be determined by optimization of the system free energy functional. 

Since SGA serves as the leading order of a systematic real-space diagrammatic expansion \cite{KaczmarczykNewJPhys2014Suppl}, additional conditions on the $\lambda$-parameters need to be imposed in order to eliminate tadpole graphs and ensure rapid convergence of the underlying real-space perturbative series \cite{BunemannEPL2012Suppl}. In turn, this allows us to substantially simplify the analysis by retaining only a subset of most relevant diagrams in this expansion (those surviving in the large lattice-coordination limit), which is done within the SGA. The constraints are:

\begin{align}
  \label{eq:constraints}
  \langle \Psi_0| \hat{P}_i^2 |\Psi_0\rangle & = 1, \\
  \langle \Psi_0|\hat{P}_i \hat{n}_{i\uparrow} \hat{P}_i |\Psi_0\rangle & = \langle \Psi_0|\hat{n}_{i\uparrow} |\Psi_0\rangle, \\
  \langle \Psi_0|\hat{P}_i \hat{n}_{i\downarrow} \hat{P}_i |\Psi_0\rangle & = \langle \Psi_0|\hat{n}_{i\downarrow} |\Psi_0\rangle.
\end{align}

\noindent
By making use of the above three conditions, only single free variational parameter per Cu site remains (without loss of generality, one can select $\lambda_{i, d}$). The other three are expressed as follows

\begin{align}
  \lambda_{i, 0} &= \sqrt{ \frac{1 - 2\lambda_{i, \uparrow}^2 (n_{i, 0} - d_{i, 0}) - \lambda_{i, d}^2 d_{i, 0}}{1 - 2 n_{i, 0} + d_{i, 0}} }  \\
  \lambda_{i, \uparrow} &= \lambda_{i, \downarrow} = \sqrt{ (n_{i, 0} - \lambda_{i, d}^2 d_{i, 0})/(n_{i, 0} - d_{i, 0}) },
\end{align}

\noindent
where $n_{i, 0} \equiv \langle \hat{n}_{i\uparrow}\rangle_0 = \langle \hat{n}_{i\downarrow}\rangle_0$ and $d_{i, 0} \equiv \langle \hat{n}_{i\uparrow}\rangle_0 \langle \hat{n}_{i\downarrow}\rangle_0 + \langle \hat{c}_{i\downarrow} \hat{c}_{i\uparrow}\rangle_0 \langle \hat{c}^\dagger_{i\uparrow} \hat{c}^\dagger_{i\downarrow}\rangle_0$ are local expectation values evaluated with the uncorrelated wave function, $|\Psi_0\rangle$. In the above, we have implicitly assumed paramagnetic state symmetry, which implies $\langle\hat{n}_{i\uparrow}\rangle_0 = \langle \hat{n}_{i\downarrow}\rangle_0$ and $\lambda_{i, \uparrow} = \lambda_{i, \downarrow}$. We emphasize that the on-site pairing amplitude $\langle \hat{c}_{i\downarrow} \hat{c}_{i\uparrow}\rangle_0$ enters the above quantities through the uncorrelated double site-occupancy, $d_{i, 0}$, and thus has observable impact on physical quantities.

For given microscopic Hamiltonian, $\hat{\mathcal{H}}$, one can derive the ground-state energy functional as $E_\mathrm{var} \equiv E_\mathrm{var}(\{P_\gamma\}, \{\lambda_\gamma\})$ using Wick's theorem. Here, $\{P_\gamma\}$ represents the set of all two-point expectation values $\langle \hat{c}^\dagger_{i\sigma} \hat{c}_{j\sigma^\prime} \rangle$ (normal) and $\langle \hat{c}_{i\sigma} \hat{c}_{j\sigma^\prime} \rangle$ and $\langle \hat{c}^\dagger_{i\sigma} \hat{c}^\dagger_{j\sigma^\prime} \rangle$ (anomalous), evaluated using uncorrelated wave function, $|{\Psi_0}\rangle$. We introduce a convenient symbolic notation $P_\gamma \equiv \langle{\Psi_0}| \hat{P}_\gamma |\Psi_0\rangle$, where $\hat{P}_\gamma$ is a fermion bilinear corresponding to the expectation value $P_\gamma$.

Plain variational wave function technique is, in principle, defined at zero temperature, where the concept of statistical ensemble notion is not needed. At $T = 0$, one needs simply to optimize the energy $E_\mathrm{var}(\{P_\gamma\}, \{\lambda_\gamma\})$ with respect to all parameters, with the constraint that there exists some Slater determinant $|\Psi_0\rangle$ such that $P_\gamma \equiv \langle{\Psi_0}| \hat{P} |\Psi_0\rangle$. Standard techniques may be employed to solve this optimization problem \cite{KaczmarczykNewJPhys2014Suppl}. From formal point of view, one should underline that our analysis is applicable to lattices of large size.

Here we use a finite-temperature generalization of this variational scheme, based on optimization the Gibbs free energy functional

\begin{align}
  \label{eq:fre_en_functional}
  \mathcal{F}(\{P_\gamma\}, \{\lambda_\gamma\}) = -\frac{1}{\beta} \ln \mathrm{Tr} \exp(-\beta \hat{\mathcal{H}}_\mathrm{eff}),
\end{align}

\noindent
with the effective Hamiltonian

\begin{align}
  \hat{\mathcal{H}}_\mathrm{eff} \equiv \sum_\gamma \frac{\partial E_\mathrm{var}}{\partial P_\gamma} \hat{P}_\gamma - \sum_\gamma \frac{\partial E_\mathrm{var}}{\partial P_\gamma} {P}_\gamma + E_\mathrm{var}.
\end{align}

\noindent
In the above, $\beta \equiv 1/(k_B T)$ is inverse temperature, expressed in energy units.

The system free energy is obtained by minimization of $\mathcal{F}$ over all fields, i.e.,

\begin{align}
  \label{eq:free_energy}
  F = \underset{P_\gamma\, \lambda_\gamma}{\mathrm{min}} \mathcal{F}(\{P_\gamma\}, \{\lambda_\gamma\}).
\end{align}

\noindent
One can show that, for $T = 0$ this formulation reduces to plain variational procedure, whereas at $T > 0$ it reflects the statistical mechanics of projected Fermi quasi-particles constructed on top of the variational ground state \cite{BunemannPhysRevB2003Suppl,FidrysiakJPCM2018Suppl}. Indeed, the conditions for the minimum read

\begin{align}
  \label{eq:free_energy_minimum_conditions}
  \frac{\partial \mathcal{F}}{\partial P_\gamma} =& \sum_\alpha \frac{\partial^2 E_\mathrm{var}}{\partial P_\gamma \partial P_\alpha} \left[ \langle \hat{P}_\alpha \rangle_0 - P_\alpha\right], \\
  \frac{\partial \mathcal{F}}{\partial \lambda_\gamma} = & \sum_\alpha \frac{\partial^2 E_\mathrm{var}}{\partial \lambda_\gamma \partial P_\alpha} \left[ \langle \hat{P}_\alpha \rangle_0 - P_\alpha\right] + \frac{\partial E_\mathrm{var}}{\partial \lambda_\gamma},
\end{align}

\noindent
where $\langle \hat{P} \rangle_0 = \mathrm{Tr} \hat{P} \exp(-\beta \hat{\mathcal{H}}_\mathrm{eff}) / \mathrm{Tr} \exp(-\beta \hat{\mathcal{H}}_\mathrm{eff})$ now denotes thermal expectation value within the statistical ensemble of quasiparticles governed by effective Hamiltonian. These equations are solved by $P_\gamma = \langle\hat{P}_\gamma\rangle_0$ and $\frac{\partial E_\mathrm{var}}{\partial \lambda_\gamma} = 0$, which is imposed in a self-consistent manner.

\begin{figure}
  \centering
	\includegraphics[width=1\linewidth]{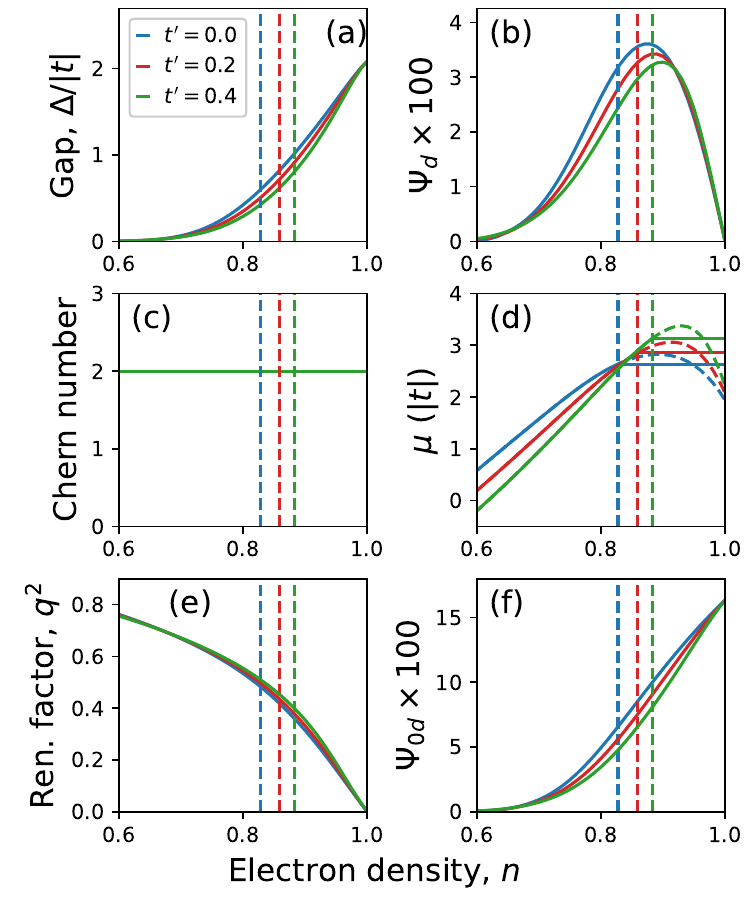}
        \caption{SGA superconducting phase diagram for the pristine triangular-lattice $t$-$J$-$U$ model ($J/|t| = 1$, $U = 16 |t|$, $k_B T = 10^{-6} |t|$) for three selections of $t^\prime/|t| = 0, 0.2$, and $0.4$. The vertical dashed lines mark the phase separation line between topological $d+id$ superconductor and Mott insulator. (a) Energy gap in the quasiparticle spectrum. (b) True $d+id$ SC order parameter amplitude (evaluated with correlated wave function). (c) Chern number. (d) Chemical potential. (e) Band narrowing factor. (f) The uncorrelated SC amplitude.}
	\label{fig:role_of_tp}
\end{figure}

The free energy functional might be recast in an alternative form

\begin{align}
  \label{eq:free_energy_and_entropy}
  F = E_\mathrm{var} - T S,
\end{align}

\noindent
where

\begin{align}
  \label{eq:entropy}
  S = - k_B \sum_\alpha \Big[\bar{n}_\alpha \ln \bar{n}_\alpha +  (1- \bar{n}_\alpha) \ln (1 - \bar{n}_\alpha) \Big],
\end{align}

\noindent
denotes system entropy. In Eq.~\eqref{eq:entropy} summation is carried out over all single-particle states of the effective Hamiltonian $\hat{\mathcal{H}}_\mathrm{eff}$ and $\bar{n}_\alpha$ are particle occupation numbers evaluated at equilibrium. From Eq.~\eqref{eq:free_energy_and_entropy} it apparent that SGA approach incorporates thermal fluctuations and may be used to study correlated superconducting state for $T > 0$. All results presented in the main text represent fully relaxed equilibrium SGA quantities, and the energy gaps ($\Delta$) are obtained by analysis of the spectrum of the effective quasiparticle Hamiltonian, $\hat{\mathcal{H}}_\mathrm{eff}$. 

\subsection{The role of next-nearest neighbor hopping}

For the sake of completeness, we incorporate in our discussion the effect of next-nearest-neighbor hopping $t^\prime$. The ratio of distances between the second-next-nearest- and nearest-neighbors in triangular lattice is $\sqrt{3}$, hence the value of $t^\prime$ is expected to be non-negligible relative to $t$, even though its absolute magnitude may be small due to overall small electronic bandwidth. In Fig.~\ref{fig:role_of_tp} we compare the SC phase diagrams obtained for the $t$-$J$-$U$ model with $t^\prime/|t| = 0, 0.2$, and $0.4$. Calculations have been carried out for $J/|t| = 1$, $U = 16 |t|$, and $k_B T = 10^{-6} |t|$.

By inspecting Fig.~\ref{fig:role_of_tp} we note that increasing $t^\prime/|t|$ results in stabilization of the homogeneous superconducting state, i.e., the phase separation boundary is shifted toward the half-filling. The gap magnitude and other parameters characterizing SC phase are not substantially affected by variation of $t^\prime$. In this respect, including $t^\prime$ appears to have similar effect on the phase separation as long-range Coulomb interactions than might be also incorporated in the $t$-$J$ or $t$-$J$-$U$ Hamiltonians \cite{WangPhysLettA2010Suppl}.

\end{document}